\documentclass[11pt]{article}

\usepackage[utf8]{inputenc}
\usepackage[T1]{fontenc}

\usepackage[english]{babel}

\usepackage{microtype}

\usepackage[hyphens]{url}
\usepackage{breakurl}

\usepackage{amsmath,amsthm,amssymb,stmaryrd,thmtools,upgreek}
\newtheorem{theorem}{Theorem}
\newtheorem{lemma}[theorem]{Lemma}

\newtheorem{conjecture}[theorem]{Conjecture}

\theoremstyle{definition}
\newtheorem{definition}[theorem]{Definition}

\newtheorem{example}[theorem]{Example}

\usepackage{graphicx}
\usepackage[obeyclassoptions,mode=tex]{standalone}
\usepackage{tikz}
\usetikzlibrary{backgrounds}
\usetikzlibrary{shapes.geometric}
\usetikzlibrary{positioning}
\usetikzlibrary{automata}
\usetikzlibrary{tikzmark}
\usetikzlibrary{patterns}
\usetikzlibrary{arrows}
\tikzset{every state/.style={minimum size=1pt}}
\usepackage{tikz-cd}

\usepackage[ruled,linesnumbered]{algorithm2e}

\usepackage{hyperref}
\usepackage[capitalise,noabbrev,nameinlink]{cleveref}
\usepackage[paper,hyperref,xcolor,cleveref]{knowledge}
\knowledgeconfigure{notion}

\crefname{conjecture}{Conjecture}{Conjectures}

\usepackage{booktabs}
\usepackage{varwidth}

\usepackage{xparse}
\usepackage{xpatch}
\usepackage{tokcycle}
\usepackage{ifthen}

\usepackage{ensps-colorscheme}

\usepackage[misc]{ifsym}
\ifXeTeX
    \usepackage{academicons}
\fi

\makeatletter
\newcommand\mathgr[1]{\tokcycle
  {\addcytoks{##1}}
  {\processtoks{##1}}
  {\ifcsname up\expandafter\@gobble\string##1\endcsname
   \addcytoks[1]{\csname up\expandafter\@gobble\string##1\endcsname}%
    \else\addcytoks{##1}\fi}
  {\addcytoks{##1}}{#1}%
  \expandafter\mathrm\expandafter{\the\cytoks}%
}
\makeatother

\NewDocumentEnvironment{proofof}{ m O{appendix} }{
    \ifcsname #1\endcsname
        \def\isInsideRestatedTheorem{1}
        \csname #1\endcsname*
    \fi
    \begin{proof}[Proof of {\cref{#1}} page {\pageref{#1}}]
        \phantomsection
        \label{#1:proof}
}{
        \ifthenelse{\equal{#2}{appendix}}{

        \noindent\hyperref[#1]{$\triangleright$ Go back to \cref{#1} on page \pageref{#1}}
        }{}
    \end{proof}
}

\NewDocumentCommand{\proofref}{ m }{
    \IfRefUndefinedExpandable{#1:proof}{}{
        \ifdefined\isInsideRestatedTheorem
        \else
        \hfill\hyperref[#1:proof]{\textsf{(Go to proof p.\pageref{#1:proof})}}
        \fi
    }
}

\NewDocumentCommand{\NewDocumentOrdering}{ m m m }{
    \expandafter\newcommand\csname #1leq\endcsname{
        \mathrel{\kl[#1]{#2}}
    }
    \expandafter\newcommand\csname #1lt\endcsname{
        \mathrel{\kl[#1]{#3}}
    }
    \knowledge{#1}{notion}
}

\NewDocumentCommand{\set}{ m }{\{ #1 \}}
\NewDocumentCommand{\setof}{ m m }{\{ #1 \mid #2 \}}
\NewDocumentCommand{\card}{ m }{\left| #1 \right|}
\NewDocumentCommand{\seqof}{ m O{n \in \Nat} }{\left( #1 \right)_{#2}}
\NewDocumentCommand{\defined}{ }{\triangleq}
\NewDocumentCommand{\range}{ O{1} m }{[#1, #2]}

\NewDocumentCommand{\Nat}{ }{\mathbb{N}}
\NewDocumentCommand{\Rel}{}{\mathbb{Z}}

\NewDocumentCommand{\bigO}{ O{} m }{\mathcal{O}_{#1}(#2)}

\NewDocumentCommand{\FO}{ }{\ensuremath{\mathsf{FO}}}
\NewDocumentCommand{\MSO}{ }{\ensuremath{\mathsf{MSO}}}

\newcommand{\ind}[1]{\mathbf{1}_{#1}}

\NewDocumentCommand{\upset}{ O{} m }{{\uparrow_{#1} #2}}
\NewDocumentCommand{\dwset}{ O{} m }{{\downarrow_{#1} #2}}

\NewDocumentCommand{\factorial}{ O{} m }{
    \if\relax\detokenize{#1}\relax
        #2!
    \else
        (#2)!
    \fi
}
\NewDocumentCommand{\app}{ m m }{\mathop{{#2} \mathrel{\kl[\app]{\triangleright}} {#1}}}
\knowledge{\app}{notion}

\NewDocumentCommand{\Res}{}{ \mathop{\kl[\Res]{\mathsf{Res}}}}
\knowledge{\Res}{notion}

\NewDocumentCommand{\resequiv}{ m m }{\mathrel{\kl[\resequiv]{\equiv_{{#1},{#2}}}}}
\knowledge{\resequiv}{notion}

\NewDocumentCommand{\resleq}{ m m }{\mathrel{\kl[\resleq]{\leq_{{#1},{#2}}}}}
\knowledge{\resleq}{notion}
\NewDocumentCommand{\resleqsf}{ m m }{\mathrel{\kl[\resleqsf]{\leq_{{#1},{#2}}^{\mathsf{sf}}}}}
\knowledge{\resleqsf}{notion}

\NewDocumentCommand{\prefleq}{}{\mathrel{\kl[\prefleq]{\sqsubseteq_{\mathsf{pref}}}}}
\knowledge{\prefleq}{notion}
\NewDocumentCommand{\prefle}{}{\sqsubset_{\mathsf{pref}}}

\NewDocumentCommand{\aTransd}{}{\mathcal{A}}

\NewDocumentCommand{\NPoly}{O{}}{\kl[\NPoly]{\mathbb{N}\mathsf{Poly}_{#1}}}
\NewDocumentCommand{\ZPoly}{O{}}{\kl[\ZPoly]{\mathbb{Z}\mathsf{Poly}_{#1}}}
\NewDocumentCommand{\NSF}{O{}}{\kl[\NSF]{\mathbb{N}\mathsf{SF}_{#1}}}
\NewDocumentCommand{\ZSF}{O{}}{\kl[\ZSF]{\mathbb{Z}\mathsf{SF}_{#1}}}
\NewDocumentCommand{\ZRat}{}{\kl[\ZRat]{\mathbb{Z}\mathsf{Series}}}
\NewDocumentCommand{\NRat}{}{\kl[\NRat]{\mathbb{N}\mathsf{Series}}}
\NewDocumentCommand{\ZCommut}{}{\kl[\ZCommut]{\mathsf{Commut}}}

\knowledge{\Poly}{notion}
\knowledge{\SF}{notion}
\knowledge{\NPoly}{notion}
\knowledge{\ZPoly}{notion}
\knowledge{\NSF}{notion}
\knowledge{\ZSF}{notion}
\knowledge{\ZRat}{notion}
\knowledge{\NRat}{notion}
\knowledge{\ZCommut}{notion}

\NewDocumentCommand{\npolyleq}{ O{} }{\mathrel{\kl[\npolyleq]{\preceq_{\Nat #1}}}}
\knowledge{\npolyleq}{notion}
\NewDocumentCommand{\zpolyequiv}{ O{} }{\mathrel{\kl[\zpolyequiv]{\equiv_{\Rel #1}}}}
\knowledge{\zpolyequiv}{notion}

\NewDocumentCommand{\BadExOk}{}{\mathsf{f}}
\NewDocumentCommand{\BadExKo}{}{\mathsf{g}}

\NewDocumentCommand{\translate}{m}{\mathop{\kl[\translate]{\tau_{#1}}}}
\knowledge{\translate}{notion}
\NewDocumentCommand{\Diff}{m m}{ \mathop{\kl[\Diff]{\Delta_{#1}}}(#2) }
\knowledge{\Diff}{notion}

\NewDocumentCommand{\vcount}{ O{} m }{|#2|_{#1}}

\newcommand{\CoveredPoly}{\kl[\CoveredPoly]{\mathsf{PolyNNeg}}}
\newcommand{\CorrectPoly}{\kl[\CorrectPoly]{\mathsf{PolyRec}}}

\NewDocumentCommand{\Deriv}{ m m m }{\mathop{\withkl{\kl[\Deriv]}{\cmdkl{\Delta}_{#3}^{#2}\mathopen{\cmdkl{[}}#1\mathclose{\cmdkl{]}}}}}
\knowledge{\Deriv}{notion}

\NewDocumentCommand{\counting}{ m }{\mathop{\withkl{\kl[\counting]}{\cmdkl{\text{nbr}[}#1\cmdkl{]}}}}
\knowledge{\counting}{notion}
\knowledge{notion}
 | commutativity
 | commutative

\knowledge{notion}
 | well

\knowledge{notion}
 | well-quasi-ordering
 | well-quasi-orderings
 | well-quasi-ordered

\knowledge{notion}
 | good sequences
 | good sequence
 | good

\knowledge{notion}
 | bad
 | bad sequence

\knowledge{notion}
 | residual
 | residuals

\knowledge{notion}
 | growth
 | growth rate

\knowledge{notion}
 | continuous

\knowledge{notion}
 | star-free 
 | star-free $\Nat $-polyregular function
 | star-free $\Rel $-polyregular function

\knowledge{notion}
 | $\Rel $-polyregular functions
 | $\Rel $-polyregular function
 | $\Rel $-polyregular
 | $\mathbb {S}$-polyregular
 | $\mathbb {S}$-polyregular functions
 | $\mathbb {S}$-polyregular function
 | $\Nat $-polyregular functions
 | $\Nat $-polyregular function
 | $\Nat $-polyregular

\knowledge{notion}
 | polyregular functions

\knowledge{notion}
 | prefix ordering

\knowledge{notion}
 | downwards closed

\knowledge{notion}
 | ultimately polynomial

\knowledge{notion}
 | noncommutative derivative

\knowledge{notion}
 | computes
 | computed

\knowledge{notion}
 | $\NPoly [0]$-transducer
 | $\NPoly [0]$-transducers
 | $\NPoly [k-1]$-transducer
 | $\NPoly [1]$-transducer
 | $\NSF [0]$-transducer
 | $\mathcal {H}$-transducer
 | $\mathcal {H}$-transducers

\knowledge{notion}
 | aperiodic monoid
 | aperiodic@monoid

\knowledge{notion}
 | residual transducer
 | residual transducers
 | $k$-residual transducers
 | $k$-residual transducer
 | $0$-residual transducer
 | $1$-residual transducer

\knowledge{notion}
 | aperiodic ordering
 | aperiodic@ordering

\knowledge{notion}
 | counter-free
\knowledge{notion}
 | counter
 | counters

\title{N-polyregular functions arise from well-quasi-orderings}
\author{%
        Aliaume Lopez\thanks{University of Warsaw}%
    ~~\href{mailto:ad.lopez@uw.edu.pl}{\Letter}%
    }

\newcommand{\makeabstract}{
\begin{abstract}
    A fundamental construction in formal language theory is the
    Myhill-Nerode congruence on words, whose finitedness characterizes
    regular language. This construction was generalized to functions
    from \(\Sigma^*\) to \(\mathbb{Z}\) by Colcombet, Douéneau-Tabot,
    and Lopez to characterize the class of so-called
    \(\mathbb{Z}\)-polyregular functions. In this paper, we relax the
    notion of equivalence relation to quasi-ordering in order to study
    the class of \(\mathbb{N}\)-polyregular functions, that plays the
    role of \(\mathbb{Z}\)-polyregular functions among functions from
    \(\Sigma^*\) to \(\mathbb{N}\). The analogue of having a finite
    index is then being a well-quasi-ordering. This provides a canonical
    object to describe \(\mathbb{N}\)-polyregular functions, together
    with a powerful new characterization of this class.
\end{abstract}
}

\begin{document}
\maketitle
\makeabstract

\section{Introduction}
\label{intro:sec}

The generalization of regular languages and their numerous equivalent
descriptions to string-to-string functions has been a central topic in automata
theory for the last decades. Generalizing the notion of two-way transducer by
adding a stack of pebbles, Engelfriet and Maneth introduced in \cite{ENMA02}
the notion of \emph{pebble transducer}, a model of computation that is able to
compute string-to-string functions with polynomial size outputs. This sparkled
the study of so-called \intro{polyregular functions} that find equivalent
characterizations notably in terms of \emph{MSO formulas} and \emph{pebble
transducers}, and \emph{simple for programs on strings} \cite{BOJA18}. One
particular restriction of polyregular functions has gotten a lot of attention
in the literature: the unary output polyregular functions. These were a central
topic of the work of Douéneau-Tabot in his Ph.D. thesis \cite[Part II]{DOUE23},
proving optimization results for pebble transducers \cite{DOUE21,DOUE22} that
fail in the general case \cite[Theorem III.3]{BOJA23}.

Unary output polyregular functions enjoy a rich algebraic structure as a
subclass of what is more well-known as ($\Nat$) \emph{(noncommutative) rational
series}, that is functions from $\Sigma^*$ to $\Nat$ computed by
$\Nat$-weigthed automata \cite{BERE88,BERE10}. In this setting,
\emph{polyregular functions} are exactly \emph{rational series} with polynomial
size outputs \cite[Theorem]{DOUE23}, a model that was already investigated in
the 1960's by \cite{SCHU62}. Following a more database oriented approach, one
can understand unary output polyregular functions as a form of counting \MSO\
queries \cite{KRRC13}, that is, $\Nat$-linear combinations of functions that
count the number of valuations of a given \MSO\ formula in a word, which is
similar in spirit to \texttt{SELECT COUNT(*) FROM word WHERE ...} queries in
SQL.

Like regular languages, polyregular functions enjoy a rich algebraic structure,
and there exists a \emph{star-free} restriction of polyregular functions that
corresponds to \emph{first order definable} functions. It was with the aim of
deciding the membership problem of star-free polyregular functions inside
polyregular functions that Colcombet, Douéneau-Tabot, and Lopez introduced the
notion of $\Rel$-polyregular functions in \cite{CDTL23}. The decidability of
the membership problem for \emph{star-free} $\Rel$-polyregular functions inside
$\Rel$-polyregular functions was proven using a notion of \emph{residual
transducer}, a computation model that is powered by computing differences of
functions. Unfortunately, the decidability result could not be generalized to
unary porylegular functions precisely because of the inability to consider the
difference of two outputs \cite[Conjecture 7.61]{DOUE23}. Furthermore, it is to
this date open whether the membership problem of $\Nat$-polyregular functions
inside $\Rel$-polyregular functions is decidable \cite[Open question
5.55]{DOUE23}.

\paragraph{Contributions.} This paper extends the notion of \emph{residual
transducer} from $\Rel$-polyregular functions to $\Nat$-polyregular functions,
effectively providing a new canonical object associated to polyregular
functions with unary output. Such a result is obtained by shifting the
attention from equivalence relations of finite index --- that are the classical
tool in automata theory --- towards order relations that are
\kl{well-quasi-ordered}, which is the order counterpart of having finite index.
We also provide a deeper understanding of the notion of \emph{residual
transducer} by illustrating how it acts as an \emph{integral operator} on
functions. This allows us to present unary output polyregular functions as
repeated ``integrals'' of a constant function, similarly to how usual
polynomials are defined as repeated primitives of a constant function.

\paragraph{Outline.} In \cref{preliminaries:sec},
we recall the necessary background on polyregular functions and
well-quasi-orderings. In
\cref{noncommutative-integration:sec},
we introduce the notion of \emph{integration operator} on functions
(\cref{oracle-transducer:def}), and prove an \emph{integration
theorem} for polyregular functions (\cref{H-transducers:thm}). In
\cref{residual-transducer:sec}, we introduce
the notion of \emph{residual transducer} for $\NPoly$
(\cref{residual-transducer:def}), and prove our main
\cref{non-commutative-npoly:thm} characterizing $\NPoly[k]$ in
terms of residual transducer and well-quasi-orderings. Finally, in
\cref{aperiodic-star-free:sec},
we discuss the generalization of these results to $\NSF$.

\section{Preliminaries}
\label{preliminaries:sec}

In this paper, we use $\Nat$ and $\Rel$ to denote respectively the set of
natural numbers and the set of integers. We use $\Sigma$ and $\Gamma$ to denote
finite alphabets, and $\Sigma^*$ to denote the set of finite words over
$\Sigma$. Given $k \in \Nat$, we also write $\Sigma^{\leq k}$ for the set of
words of length at most $k$. The size of a word $w \in \Sigma^*$ is denoted
$\card{w}$, and $\card{w}_a$ is the number of occurrences of the letter $a$ in
$w$. We also allow ourselves to write $w_{\leq i}$ for the (possibly empty)
prefix of $w$ of length $i$, and $w_{> i}$ for the (possibly empty) suffix of
$w$ starting at position $i+1$. We use the symbol $\varepsilon$ for the empty
word. In general, we assume that the reader is familiar with the basics of
automata theory and formal languages such as finite state automata, Monadic
Second Order Logic (\MSO) and First Order Logic (\FO) on words, and refer to
\cite{THOM97} for a comprehensive introduction.

Let us first recall the definition of a \emph{polyregular function} in the
commutative output case, that is the definition of respectively
$\Nat$-polyregular functions and $\Rel$-polyregular functions. For simplicity,
we provide the definition of $\mathbb{S}$-polyregular functions for an
arbitrary commutative semiring $\mathbb{S}$.

\begin{definition}
    \label{polyregular-function:def}
    Let $\mathbb{S}$ be a commutative semiring.
    A function $f \colon \Sigma^* \to \mathbb{S}$ is \intro{$\mathbb{S}$-polyregular} if
    there exists a finite set of $\MSO$-formulas $\varphi_1, \ldots, \varphi_n$
    with first-order free variables $\vec{x}$, and a finite set of
    constants $c_1, \ldots, c_n \in \mathbb{S}$, such that
    for all $w \in \Sigma^*$:
    \begin{equation*}
        f(w) = \sum_{i=1}^n c_i \cdot \counting{\varphi}(w)
    \end{equation*}
    Where $\intro*\counting{\varphi_i}(w)$ is the number of valuations of $\vec{x}$
    that satisfy $\varphi_i$ in the word $w$.
    The function $f$ is \intro{star-free} if the formulas $\varphi_i$ are
    in $\FO$.
\end{definition}

As an example, the function $w \mapsto \card{w}_a \times \card{w}_b$ is
computable using the formula $\varphi(x,y) \defined a(x) \land b(y)$, and is
therefore a \kl{star-free} \kl{$\Nat$-polyregular function}.

\AP Let us write $\intro*\NPoly[k]$ for the set of functions $f \colon \Sigma^*
\to \Rel$ that are computable by an $\MSO$-formula using at most $k$ free
variables, and $\intro*\ZPoly[k]$ for the set of functions $f \colon \Sigma^*
\to \Nat$ that are computable by an $\MSO$-formula using at most $k$ free
variables. The analogue notions of $\intro*\NSF[k]$ and $\intro*\ZSF[k]$ are
defined as their counterpart for \kl{star-free} functions. 

\AP Let us end this preliminary section by introducing the necessary tools on
\emph{well-quasi-orderings} that we will rely upon in the rest of this
document, and refer to \cite{SCSC12} for a more comprehensive overview of this
area of computer science. A sequence $\seqof{u_i}[i \in \Nat]$ of elements in a
quasi-ordered set $(X, \leq)$ is \intro{good} whenever there exist $i < j$ such
that $u_i \leq u_j$. The set $X$ is a \intro{well-quasi-ordering} when every
infinite sequence is \kl{good}. A sequence is \intro{bad} when it is not
\kl{good}. The notion of \kl{good sequences} can also be applied to binary
relations that are not orderings, and a binary relation $R$ for which every
infinite sequence is \kl{good} is said to be \intro{well} \cite{MELL98}.\footnote{
    We apologize for the confusing terminology.
} The
class of \kl{well-quasi-orderings} contains $(\Nat, \leq)$, and is closed under
taking finite products \cite[Dickson's lemma]{SCSC12}

\section{Noncommutative Integration Operators}
\label{noncommutative-integration:sec}

In this section, we re-introduce the notion of transducers with oracle of
\cite{CDTL23}, and re-interpret them as \emph{integration operators} on
functions. To illustrate the construction on a simple example, let us consider
a function $f \colon \set{1}^* \to \set{1}^*$, that is a function $f \colon
\Nat \to \Nat$ with numbers represented in unary.
Let us write $\Delta[f] (n) = f(n+1) - f(n)$ for the \emph{discrete derivative}
of $f$ at $n$, and let us remark that for all $n \in \Nat$,
\begin{equation}
    \label{discrete-derivative:eq}
    f(n) = f(0) + \sum_{i=0}^{n-1} \Delta[f](i)
    \quad .
\end{equation}
A presentation of \cref{discrete-derivative:eq} in terms of transducers
would lead to the following picture:
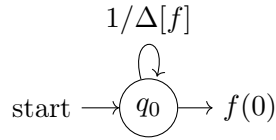
\begin{center}
    \begin{tikzpicture}
        \node[state, initial] (q0) {$q_0$};
        \node[right=1.2em of q0] (q1) {$f(0)$};
        \draw (q0) edge[loop above] node{$1/\Delta[f]$} (q0);
        \draw[->] (q0) edge (q1);

    \end{tikzpicture}
\end{center}
Where the semantics of such a model is to
start in node $q_0$, and when reading a $1$, output the value of $\Delta[f]$
on the rest of the input, and loop back to $q_0$. When the input is empty, the
automaton outputs $f(0)$.

However, when the input alphabet is not unary, the notion of $\Delta[f]$ is not
well-defined, and we need to consider a more general notion of \emph{integration operator}.
\begin{definition}
    \label{oracle-transducer:def}
    Let $\mathcal{H} \subseteq \Rel^{\Sigma^*}$ be a set of
    functions. An \intro{$\mathcal{H}$-transducer}
    is a tuple $\aTransd \defined (Q, q_0, \delta, \lambda, F)$
    where
    \begin{itemize}
        \item $Q$ is a finite set of states,
        \item $q_0 \in Q$ is called the initial state,
        \item $\delta \colon Q \times \Sigma \to Q$ is called the transition function,
        \item $\lambda \colon Q \times \Sigma \to \mathcal{H}$ is called the integrated function,
        \item $F \colon Q \to \Rel$ is called the final condition.
    \end{itemize}
    We allow ourselves to write $\delta^* \colon Q \times \Sigma^* \to Q$ for
    the repeated application of the transition function $\delta$, that is
    $\delta^*(q,\varepsilon) \defined q$, and $\delta^*(q, aw) \defined \delta(\delta^*(q,a), w)$.
\end{definition}

\AP The semantics of an \kl{$\mathcal{H}$-transducer} is defined by induction.
We say that a transducer $\aTransd$ \intro{computes} a function $f \colon
\Sigma^* \to \Rel$ if for all $q \in Q$, $a \in \Sigma$, and $w \in \Sigma^*$,
$\aTransd(q, aw) = \aTransd(\delta(q,a), w) + \lambda(q, a)(w)$ and
$\aTransd(q, \varepsilon) = F(q)$.
That is, given a word $w$:
\begin{equation}
    \label{transducer-semantics:eq}
    \aTransd(q_0, w) 
    = \sum_{i=0}^{|w|-1} \lambda(\delta^*(q_0, w_{\leq i}), w_{i+1})(w_{> i+1}) + F(\delta^*(q_0, w))
    \quad 
    .
\end{equation}

\AP In that sense, the $\mathcal{H}$-transducer is really computing a
\emph{noncommutative integral} of the function $\lambda$ with (final) condition
$F$. The analogy between \cref{transducer-semantics:eq} and
\cref{discrete-derivative:eq}, does not stop here, as the notion of
\emph{polynomials} can be extended to \emph{polyregular functions} using the
following ``integration theorem for polyregular functions.'' In order to
formally state \cref{H-transducers:thm}, we need to introduce the
automata-theoretic counterpart of \FO, which is based on the notion of
\kl{counter}: a \intro{counter} in a finite state automaton is a pair $(q,u)$
where $q$ is a state and $u$ is a word such that $\delta(q,u) \neq q$ and
$\delta(q,u^n) = q$ for some $n \geq 2$. An automaton is \intro{counter-free}
when it contains no \kl{counters}. This connection between \kl{counter-free}
automata and \kl{star-free} functions (with boolean output) is well-known in
automata theory \cite{MNPA71}.

\begin{theorem}[\cite{DOUE23}]
    \label{H-transducers:thm}
    Let $f \colon \Sigma^* \to \Rel$ (resp. $\Nat$) be a function and $k \geq 1$.
    Then,
    $f \in \NPoly[k]$
    if and only if $f$ is computed by an \kl{$\NPoly[k-1]$-transducer} 
    Furthermore, $f \in \NSF[k]$
    if and only if
    $f$ is computed by a \kl{counter-free} $\NSF[k-1]$-transducer.
    The same holds for the classes $\ZPoly[k]$ and $\ZSF[k]$.
\end{theorem}

However, \cref{H-transducers:thm} does not provide a
\emph{canonical} model for computing a function $f$, and therefore cannot help
in deciding membership, equivalence, or optimization problems. This can already
be seen in the case of unary input functions, as illustrated by the two
distinct minimal (in the number of states) \kl{$\NPoly[0]$-transducers} of
\cref{non-canonical-transd:fig}
computing
the function $\BadExOk$ of \cref{non-canonical-transd:ex}.

\begin{figure}
    \centering
    \begin{tikzpicture}
        \begin{scope}[xshift=-3cm]
            \node[state, initial] (q0) {$q_0$};
            \node[state, below=of q0] (q1) {$q_1$};
            \node[right=1.5em of q0] (o0) {$1$};
            \node[right=1.5em of q1] (o1) {$0$};
            \draw[->] (q0) edge (o0);
            \draw[->] (q1) edge (o1);
            \draw[->] (q1) edge[loop left] node{$a/1$} (q1);
            \draw[->] (q0) edge node[midway, left] {$a/0$} (q1);
        \end{scope}
        \begin{scope}[xshift=3cm]
            \node[state, initial] (p0) {$q_0$};
            \node[state, below=of p0] (p1) {$q_1$};
            \node[right=1.5em of p0] (o0) {$1$};
            \node[right=1.5em of p1] (o1) {$0$};
            \draw[->] (p0) edge (o0);
            \draw[->] (p1) edge (o1);
            \draw[->] (p0) edge[bend right=30] node[midway, left] {$a/0$} (p1);
            \draw[->] (p1) edge[bend right=30] node[midway, right] {$a/ \lambda u. 2 \times \ind{u \neq \varepsilon}$} (p0);
        \end{scope}
    \end{tikzpicture}
    \caption{Two distinct $\NPoly[0]$-transducers computing the function $\BadExOk$ of \cref
    {non-canonical-transd:ex}.
}
    \label{non-canonical-transd:fig}
\end{figure}

\begin{example}
    \label{non-canonical-transd:ex}
    Let $\BadExOk \colon \set{a}^* \to \Nat$ that maps 
    $\varepsilon$ to $1$, and $aw$ to $\card{w}$.
    Then $\BadExOk$ is computed by the two transducers of 
    \cref{non-canonical-transd:fig}, and no \kl{$\NPoly[0]$-transducer} with fewer 
    states can compute $\BadExOk$.
\end{example}
\begin{proof}
    Assume by contradiction that there exists an \kl{$\NPoly[0]$-transducer}
    with one state that computes $\BadExOk$.
    Then, the transducer must output $1$ on the empty word,
    and has a self-loop $\delta(q_0) = q_0$.
    Assuming that it computes $\BadExOk$, we must have for all $u \in \Sigma^*$,
    $f(au) = \aTransd(q_0, au) =
    \aTransd(q_0, u) + \lambda(q_0, a)(u) = 
    f(u) + \lambda(q_0, a)(u)$.
    Therefore,
    $\lambda(q_0, a)(u) = f(au) - f(u)$ takes a negative
    value for $u = \varepsilon$, which is absurd because
    $\lambda(q_0, a) \in \NPoly[0]$.

    Let us prove that the left transducer of
    \cref{non-canonical-transd:fig}
    computes $f$.
    We prove by induction on $w$
    that $A(q_0, w) = f(w)$,
    and that $A(q_1, w) = f(aw)$.
    To that end, let us first remark that $f(\varepsilon) = 1$,
    and $f(au) = \card{u}$.
    When $w = \varepsilon$, $A(q_0, \varepsilon) = F(\varepsilon) = 1 = f(\varepsilon)$.
    Similarly, $A(q_1, \varepsilon) = F(q_1) = 0 = f(a)$.

    Assume that $w = au$. Then:
    \begin{align*}
        A(q_0, w) = A(q_0, au) &= \lambda(q_0, a)(u) + A(q_1, u) \\ 
                               &= 0 + A(q_1,u) \\
                               &= f(au) & \text{by induction hypothesis} \\
                               &= f(w) 
    \end{align*}
    Similarly,
    \begin{align*}
        A(q_1, w) = A(q_1, au) &= \lambda(q_1, a)(u) + A(q_1, u) \\ 
                               &= 1 + A(q_1,u) \\
                               &= 1 + f(au) & \text{by induction hypothesis} \\
                               &= 1 + \card{u} \\
                               &= \card{au} \\
                               &= f(aau) \\
    \end{align*}

    The proof for the other automaton is similar. The
    key ingredient in the induction hypothesis is that if 
    $\card{u} \geq 1$, then:
    $f(aau) - f(u) = 2$ and otherwise $f(aau) - f(u) = 0$.
    Hence, $f(aau) - f(u) = 2 \times \ind{u \neq \varepsilon} 
    = \lambda(q_1, a)(u)$.
\end{proof}

\section{Residual Transducers}
\label{residual-transducer:sec}

\AP In order to obtain a \emph{canonical object} based on
\kl{$\mathcal{H}$-transducers}, a key ingredient is to consider the so-called
\intro{residuals} of a function $f \colon \Sigma^* \to \Rel$,  defined by
$\intro*\app{f}{u} \defined w \mapsto f(uw)$. The collection of \kl{residuals}
of a function $f$ is denoted $\intro*\Res(f)$ and is defined as the set of
$\app{f}{u}$ where $u$ ranges over words in $\Sigma^*$. Given a function $f
\colon \Sigma^* \to \Nat$ and two words $u,v \in \Sigma^*$, we also define the
\intro{noncommutative derivative} $\intro*\Deriv{f}{u}{v} \defined \app{f}{u} -
\app{f}{v}$.

\AP To compute a function $f$ by ``integrating'' simpler functions, we
will try to detect when the \kl{noncommutative derivative} of $f$ is simpler
than $f$ itself. To that end, given $k \in \Nat$, we define the following
equivalence relation on $\Sigma^*$: $u \intro*\resequiv{f}{k} v$ if and only if
$\Deriv{f}{u}{v} \in \ZPoly[k-1]$. Note that $\resequiv{f}{k}$ is an
equivalence relation because $\ZPoly[k-1]$ is closed under $\Rel$-linear
combinations and multiplication by $(-1)$. The key construction of Colcombet,
Douéneau-Tabot, and Lopez in \cite{CDTL23} is to remark that this equivalence
relation characterizes $\ZPoly[k]$ in the following sense: given $f \colon
\Sigma^* \to \Rel$, $f \in \ZPoly[k]$ if and only if $\resequiv{f}{k}$ has
finite index.

\AP The analogue notion for $\NPoly[k-1]$ is defined by $v \intro*\resleq{f}{k}
u$ if and only if $\Deriv{f}{u}{v} \in \NPoly[k-1]$, this is only a partial
ordering relation. Given a function $f$, our
goal is to leverage $\resleq{f}{k}$ to build a canonical
\kl{$\NPoly[k-1]$-transducer} that \kl{computes} $f$. The idea is to consider
as states the minimal elements of $\Sigma^*$ for $\resleq{f}{k}$, and define
transitions by letting $\delta(u, a)$ be some state $v$ such that $v
\resleq{f}{k} ua$. To produce a canonical model, this has to be done carefully,
as illustrated by the two distinct \kl{$\NPoly[0]$-transducers} of
\cref{non-canonical-transd:fig}
computing
the function $\BadExOk$ of \cref{non-canonical-transd:ex}, and
having as states minimal elements for $\resleq{\BadExOk}{0}$. To ensure
unicity, we ask that the set of states is a \intro{downwards closed} subset of
$\Sigma^*$ for the \intro{prefix ordering} $(\intro*\prefleq)$, i.e. that every
prefix of a state is also a state. 

\begin{lemma}
    \label{good-residual-ordering:fact}
    Let $k \in \Nat$, and let $f \colon \Sigma^* \to \Nat$. Then,
    $(\resleq{f}{k})$ is a quasi-ordering, satisfying the following
    extra properties:
    \begin{enumerate}
        \item For all $u,v,w \in \Sigma^*$, $u \resleq{f}{k} v$
            implies $uw \resleq{f}{k} vw$,
        \item If $u \resleq{f}{k} v$ and $v \resleq{f}{k} u$,
            then $\Deriv{f}{u}{v} = 0$,
    \end{enumerate}
\end{lemma}
\begin{proof}
    Assume that $u \resleq{f}{k} v$ and $v \resleq{f}{k} w$.
    Then, $\Deriv{f}{u}{v} \in \NPoly[k-1]$ and $\Deriv{f}{v}{w} \in \NPoly[k-1]$,
    we conclude that $\Deriv{f}{u}{w} = \Deriv{f}{u}{v} + \Deriv{f}{v}{w} \in \NPoly[k-1]$.
    We have proven that $(\resleq{f}{k})$ is a quasi-ordering.

    Let $u,v,w \in \Sigma^*$ be such that $u \resleq{f}{k} v$.
    Then, $\Deriv{f}{u}{v} \in \NPoly[k-1]$.
    We have $\Deriv{f}{uw}{vw} = \app{\Deriv{f}{u}{v}}{w} \in \NPoly[k-1]$.

    Finally, let $u,v \in \Sigma^*$ be such that $u \resleq{f}{k} v$ and $v \resleq{f}{k} u$.
    Then, $\Deriv{f}{u}{v} \in \NPoly[k-1]$ and $\Deriv{f}{v}{u} \in \NPoly[k-1]$,
    as a consequence $\app{f}{u} - \app{f}{v}$ is non-negative,
    and $\app{f}{v} - \app{f}{u}$ is non-negative, leading to 
    $\app{f}{u} = \app{f}{v}$, i.e.,
    $\Deriv{f}{u}{v}  = \Deriv{f}{v}{u} = 0$.
\end{proof}

\begin{definition}
    \label{residual-transducer:def}
    Let $f \colon \Sigma^* \to \Nat$ and $k \in \Nat$.
    A transducer $\aTransd \defined (Q, q_0, \delta, \lambda, F)$
    is a \intro{$k$-residual transducer}
    of $f$ 
    when
    it is a \kl{$\NPoly[k-1]$-transducer}
    satisfying the following properties:
    \begin{enumerate}
        \item $\aTransd$ \kl{computes} $f$;
        \item $Q \subseteq \Sigma^*$ is a \kl{downwards closed}
            for $\prefleq$;
        \item $q_0 = \varepsilon$;
        \item every state $q \in Q$ is accessible from $q_0$;
        \item For all $u, a \in Q$,
            $\delta(u,a)$ is the $\prefleq$-maximal $v \in Q$
            such that $v \prefleq ua$, and $v \resleq{f}{k} ua$.
        \item For all $u,a \in Q$,
            $\lambda(u,a) = 
            \Deriv{f}{ua}{\delta(u,a)} \in \NPoly[k-1]$.
    \end{enumerate}
\end{definition}

Let us immediately prove that \kl{$k$-residual transducers} are unique due to
the way their state space  is defined and transitions are computed using
maximal prefixes.

\begin{lemma}
    \label{unique-res-transducer:fact}
    Let $f \colon \Sigma^* \to \Nat$ and $k \in \Nat$.
    Then $f$ has at most one \kl{$k$-residual transducer}.
\end{lemma}
\begin{proof}
    Let $\aTransd_1$ and $\aTransd_2$ be two
    \kl{$k$-residual transducers} for $f$.
    The two initial states must be $\varepsilon$.
    Let us prove by induction on $u \in \Sigma^*$ that
    $\delta_1(\varepsilon, u) = \delta_2(\varepsilon, u)$
    and that $Q_1$ equals $Q_2$ over prefixes of $u$.
    This will prove that 
    $Q_1 = Q_2$, hence that $\aTransd_1 = \aTransd_2$.

    Let $u \in \Sigma^* \cap Q_1 \cap Q_2$ and $a \in \Sigma$, $v_1 \in Q_1$ be
    defined as $v \defined \delta_1(u,a)$, and $v_2 \defined \delta_2(u,a)$.
    Remark that by induction hypothesis, for all $v \prefle u$, $v \in Q_1 \cap
    Q_2$. If $\delta_1(u,a) = Q_1$, it means that for all $v \in Q_2$ such that
    $v \prefle ua$, we have $\neg( v \resleq{f}{k} ua )$. The only possible
    transition in $\aTransd_2$ is therefore $\delta_2(u,a) = ua$, and $ua \in
    Q_2$. Similarly, if $\delta_1(u,a) \prefleq u$, then $\delta_2(u,a) =
    \delta_1(u,a)$ by definition of $\delta_2$ as a maximum.
\end{proof}

Let us now introduce \cref{residual:algo} to compute the \kl{$k$-residual
transducer} given a function $f$. Notice that this algorithm requires the
ability to test if a function belongs to $\NPoly[k]$, which is only known to be
feasible for \kl{commutative} \kl{polyregular functions}
\cite[Theorem 31]{LOPEZ24}.
However, the
termination of this algorithm also proves the existence of the \kl{$k$-residual
transducer}. The key argument proving the termination of \cref{residual:algo}
is based on the fact that for a function $f \in \NPoly[k]$, the quasi-ordering
$(\resleq{f}{k})$ is a \kl{well-quasi-ordering}. This connection is made
possible through the characterization of $\NPoly$ as counting the number of
valuations of some $\MSO$ formulas (see \cref{polyregular-function:def}).

\begin{algorithm}[t]
    $Q \defined \{ \varepsilon \}$;

    $O \defined \setof{ a }{ a \in \Sigma}$;

    $\delta \defined \emptyset$;

    $\lambda \defined \emptyset$;

    $F \defined \emptyset$;

    \While{$O \neq \emptyset$}{
        choose $ua \in O$;

        $O \defined O \setminus \set{ ua }$;

        \eIf{$\exists v \in Q, v = \max_{\prefleq} \setof{w \in Q}{w \prefleq ua \wedge w \resleq{f}{k} ua}$}{
            $\delta(u, a) \defined v$;

            $\lambda(u, a) \defined \Deriv{f}{ua}{v}$;
        }{
            $Q \defined Q \uplus \set{ ua }$;

            $\delta(u,a) \defined ua$;

            $\lambda(u,a) \defined 0$;

            $O \defined O \cup \setof{ uab }{b \in \Sigma}$;
        }
    }
    \For{$u \in Q$}{
        $F(u) \defined f(u)$;
    }
    \Return{$(Q, \varepsilon, \delta, \lambda, F)$};
    \caption{Computing a $k$-residual transducer of a function $f$.}
    \label{residual:algo}
\end{algorithm}

\paragraph{Correctness of the Algorithm.} Let us now prove that
\cref{residual:algo}
computes
the \kl{$k$-residual transducer} of a function $f$. We start by collecting some
invariants about its execution in \cref{q-o-prefix-cool:fact},
allowing us to derive correctness in \cref{correct-residual:lemma}.
The termination of the algorithm is then obtained in
\cref{wqo-implies-termination:lemma}, under the assumption that
$\resleq{f}{k}$ is a \kl{well-quasi-ordering}.

\begin{lemma}
    \label{q-o-prefix-cool:fact}
    Let $f \colon \Sigma^* \to \Nat$ and $k \in \Nat$.
    At each step of the \texttt{while loop}
    of \cref{residual:algo}, the sets
    $Q$ and $O$ are such that
    \begin{enumerate}
        \item $Q \cup O$ is a \kl{downwards closed} subset of 
            $\Sigma^*$ for $\prefleq$;
        \item elements in $O$ are pairwise incomparable
            for $\prefleq$, and are maximal
            for $\prefleq$ inside $Q \cup O$.
    \end{enumerate}
\end{lemma}
\begin{proof}
    Let us write $Q_i$ and $O_i$ for the value of the variables
    $Q$ and $O$ at step $i$ of the \texttt{while loop}.
    We prove the desired property by induction on $i$.

    For $i=0$, the property is true because
    $Q_0 = \set{\varepsilon}$ and $O_0 = \setof{a}{a \in \Sigma}$.

    For $i+1$. Either the \texttt{if} branch was taken, in which case $Q_{i+1}
    \cup O_{i+1} = (Q_i \cup O_i) \setminus \set{u}$ for some $u \in O_i$. This
    set remains \kl{downwards closed}, and elements in $O_{i+1}$ remain maximal
    elements. 

    If the \texttt{else} branch was taken, then there exists $u \in O_i$ such
    that $Q_{i+1} = Q_i \cup \set{u}$ and $O_{i+1} = O_i \setminus \set{ u }
    \cup \setof{ ua }{ a \in \Sigma}$. We conclude that $Q_{i+1} \cup O_{i+1} =
    Q_i \cup O_i \cup \setof{ ua }{a \in \Sigma}$ continues to be \kl{downwards
    closed} for $\prefleq$. Let $v \in Q_{i+1} \cup O_{i+1}$ be such that $ua
    \prefleq v$ for some $a \in \Sigma$. Then $u \prefleq v$, and $u = v$ since
    $u$ was a maximal element. As a consequence, $ua$ is a maximal element for
    all $a \in \Sigma$. Assume by contradiction that $ua$ is comparable with
    some $v \in O_{i+1}$ with $ua \neq v$, it cannot be that $ua \prefleq v$ by
    the above argument, and if $v \prefleq ua$ with $v \neq ua$, then $v
    \prefleq u$ and $u = v$, which is absurd since $v \not \in O_{i+1}$.
    We have concluded that $O_{i+1}$ continues to have pairwise incomparable
    elements.
\end{proof}

\begin{lemma}
    \label{correct-residual:lemma}
    If \cref{residual:algo} terminates on 
    an input $f \colon \Sigma^* \to \Nat$, $k \in \Nat$,
    then it computes the \kl{$k$-residual transducer} of $f$.
\end{lemma}
\begin{proof}
    Because of \cref{q-o-prefix-cool:fact},
    we already know that $q_0 = \varepsilon$,
    $Q$ is a \kl{downwards closed} subset of $\Sigma^*$
    for $\prefleq$, 
    that every state of $Q$ is accessible from $q_0$.
    Notice that at every step,
    $\lambda(u,a)$ is defined as
    $\Deriv{f}{ua}{\delta(u,a)} = \app{f}{ua} - \app{f}{\delta(u,a)}$.
    Finally, since $Q \cup O$ is a \kl{downwards closed} subset of $\Sigma^*$
    at every step,
    we have that at step $i$,
    for all $ua \in O_i$,
    $\setof{w \in Q}{w \prefleq ua} = \setof{w \in Q_i}{ w \prefleq ua}$,
    which proves that the maximum considered in the algorithm
    is indeed computing correctly.

    Let us now prove that $\aTransd$ \kl{computes} $f$.
    To that end, let us prove by induction on $w \in \Sigma^*$ 
    that for all $q \in Q$, $\aTransd(q, w) = f(qw)$.
    The base case is trivial, as $\aTransd(q, \varepsilon) = F(q) \defined f(q)$.
    For the induction step, 
    let $w = au$ with $a \in \Sigma$ and $u \in \Sigma^*$.
    Then $\aTransd(q, w) = \aTransd(\delta(q,a), u) + \lambda(q,a)(u)$.
    By induction hypothesis, $\aTransd(\delta(q,a), u) = f(\delta(q,a)u)$.
    Furthermore, $\lambda(q,a)(u) = \Deriv{f}{qa}{\delta(q,a)} = f(qau) - f(\delta(q,a)u)$.
    As a consequence, $\aTransd(q, w) = f(qw)$ which concludes the proof.
\end{proof}

\begin{lemma}
    \label{wqo-implies-termination:lemma}
    Let $f \colon \Sigma^* \to \Nat$, and $k \in \Nat$ be such that
    every infinite, $\prefleq$-increasing sequence is \kl{good}
    in $(\Sigma^*, \resleq{f}{k})$
    (or equivalently, such that the relation $(\prefleq \Rightarrow \resleq{f}{k})$
    is \kl{well}).
    Then, \cref{residual:algo} terminates on the input $(f,k)$.
\end{lemma}
\begin{proof}
    Assume towards a contradiction that
    \cref{residual:algo} does not terminate.
    Then, the \texttt{else} branch in the \texttt{while loop}
    must be taken infinitely often.
    This means that the set $Q$ of states grows arbitrarily large.

    Let us write $\seqof{Q_i}[i \in \Nat]$ for the set of states $Q$ at step
    $i$ of the execution of \cref{residual:algo}. Applying
    \cref{q-o-prefix-cool:fact}, we know that for all $i \in \Nat$, $Q_i$ is
    \kl{downwards closed} for $\prefleq$. Let us write $Q_\infty \defined
    \bigcup_{i \in \Nat} Q_i$. The set $Q_\infty$ is infinite, and is
    \kl{downwards closed} for $\prefleq$. As a consequence, it is an infinite
    tree with a finite branching (at most $\card{\Sigma}$), and has an infinite
    branch $\seqof{u_j}[j \in \Nat]$ thanks to König's lemma.

    Let us prove that this infinite branch is a \kl{bad sequence} for the
    ordering $\resleq{f}{k}$.
    Let $j < p$, and assume by contradiction that $u_j \resleq{f}{k} u_p$. We
    know that $u_j \in Q_j$ and $u_p \in Q_p$. Then, at step $p-1$ of the
    algorithm, $u_j \in Q_{p-1}$, since $u_j \in Q_j \subseteq Q_{p-1}$.
    Because $u_j \prefleq u_p$ and $u_j \resleq{f}{k} u_p$,
    \cref{residual:algo} must take the \texttt{if} branch at step $p-1$. As a
    consequence, $u_p \not\in Q_{p}$, which is absurd.

    We have proven that the infinite branch is a \kl{bad sequence}
    for $\resleq{f}{k}$, which contradicts the assumption.
    Hence, \cref{residual:algo} must terminate.
\end{proof}

\paragraph{Validity of the algortihm.}
Let us now prove that \cref{residual:algo} is correct terminates when $f \in \NPoly[k]$, by proving that 
$\resleq{f}{k}$ is a \kl{well-quasi-ordering}, which 
is the  content of \cref{n-poly-k-implies-wqo:lemma}.

\begin{lemma}
    \label{n-poly-k-implies-wqo:lemma}
    Let $k \in \Nat$, and let $f \in \NPoly[k]$.
    Then, the relation $\resleq{f}{k}$ is a \kl{well-quasi-ordering}
    for $\Sigma^*$.
\end{lemma}
\begin{proof}
    Because $f \in \NPoly[k]$, there exists
    a tuple $\vec{x}$ of first order free variables,
    $\MSO$ formulas $\seqof{\psi_i(\vec{x})}[1 \leq i \leq n]$,
    and non-negative coefficients $\seqof{m_i}[1 \leq i \leq n]$,
    such that
    $f = \sum_{i = 1}^n m_i \times \counting{\varphi_i(\vec{x})}$.

    Let $q$ be the maximal quantifier rank of formulas $\seqof{\psi_i}[1 \leq i
    \leq n]$. To a word $u \in \Sigma^*$, we associate the vector $\MSO^q(u)$
    that maps an $\MSO$-type with $\ell \leq |\vec{x}|$ free variables to the
    number of realizations of this type in $u$.

    Let $u, v \in \Sigma^*$ such that $\MSO^q(u) \leq \MSO^q(v)$, which means
    that every $\MSO$ type (of quantifier rank $q$ and with at most $n$ free
    variables) has at least as many realizations in $v$ than it has in $u$.
    Remark that by the compositionality of $\MSO$ over words (this is an instance
    of the more general
    the Feferman-Vaught-Mostowski theorem \cite{FEVAU59,MAKOW04}), for all $\MSO^q$ types
    $t(\vec{x})$, there are finitely many $\MSO^q$ types $t_l^j(\vec{y_i}),
    t_r^j(\vec{z_i})$ with $\vec{x} = \vec{y_i} \uplus \vec{z_i}$
    for $1 \leq j \leq N_0$, such that for every
    tuple $\vec{a}$ of elements in a word $uv$, $\MSO^q(\vec{a} / uv) =
    t(\vec{x})$ if and only if there exists $1 \leq j \leq N_0$,
    such that $\vec{a} = \vec{b} \uplus \vec{c}$,
    $\MSO^q(\vec{b} / u) =
    t_l^j(\vec{y_i})$, and $\MSO^q(\vec{c} / v) = t_r^j(\vec{z_i})$.
    We write $t = t_l \odot t_r$ to signify
    that $\MSO^q(\vec{bc} / uv) = t$
    if and only if $\MSO^q(\vec{b}/u) = t_l$
    and $\MSO^q(\vec{c}/v) = t_r$.

    As a consequence, if $\MSO^q(u) \leq \MSO^q(v)$, then 
    for all $w \in \Sigma^*$:
    \begin{align*}
        & \Deriv{f}{v}{u}(w) \\
        &= f(vw) - f(uw) \\
        &= 
        \sum_{i = 1}^n m_i
        \left[
            \counting{\phi_i(\vec{x})} (vw) -
            \counting{\phi_i(\vec{x})} (uw)
        \right] \\
        &= 
        \sum_{i = 1}^n
        m_i
            \sum_{\phi_i \in t(\vec{x})}
        \left[
            \counting{t(\vec{x})}(vw)
            -
            \counting{t(\vec{x})}(uw)
        \right] \\
        &= 
        \sum_{i = 1}^n
        m_i
        \sum_{1 \leq j \leq N_0}
        \sum_{\phi_i \in t_l^j(\vec{y}) \odot t_r^j(\vec{z})}
        \underbrace{
        \left[
            \counting{t_r^j(\vec{y})}(v)
            -
            \counting{t_r^j(\vec{y})}(u)
        \right] 
    }_{ \in \Nat }
            \times 
            \counting{t_l^j(\vec{z})}(w)
            \\
        &\Deriv{f}{v}{u} \in \NPoly[k-1]
    \end{align*}

    We have proven that if $\MSO^q(u) \leq \MSO^q(v)$, then $u \resleq{f}{k}
    v$. Recall that $\Nat^p$ is a \kl{well-quasi-ordering} when endowed with
    the product ordering, and therefore that $\setof{\MSO^q(u)}{u \in
    \Sigma^*}$ is a \kl{well-quasi-ordering}.

    Let $\seqof{u_i}[i \in \Nat]$ be an infinite sequence of $\Sigma^*$.
    Without loss of generality, one can assume that for all $i \neq j$, $u_i
    \equiv_k u_j$, i.e., that the difference $\app{f}{u_i} - \app{f}{u_j}$
    belongs to $\ZPoly[k-1]$, since the latter has finite index. Thanks to the
    above remarks, there exists $i < j$ such that $\MSO^q(u_i) \leq
    \MSO^q(u_j)$. As a consequence, $g \defined \app{f}{u_j} - \app{f}{u_i} \in
    \NPoly$, and therefore $g \in \NPoly[k-1]$. We have proven that there
    exists $i < j$ such that $u_i \resleq{f}{k} u_j$.
\end{proof}

We now have all the ingredients to prove the main result of this paper, which
is a characterization of $\NPoly$ in terms of the existence of \kl{$k$-residual
transducers} and the \kl{well-quasi-ordering} of the relation
$(\resleq{f}{k})$. Note however that this theorem is not effective, as it
requires the ability to decide if a function belongs to $\NPoly[k]$ which is an
open problem \cite{LOPEZ24,DOUE23}.

\begin{theorem}
    \label{non-commutative-npoly:thm}
    Let $f \in \ZPoly$ be a non-negative function, 
    and $k \in \Nat$,
    the following are equivalent:
    \begin{enumerate}
        \item \label{n-poly-1-transd:item} $f$ is \kl{computed}
            by an \kl{$\NPoly[k-1]$-transducer};
        \item \label{n-poly-k:item} $f \in \NPoly[k]$;
        \item \label{n-poly-wqo:item} $(\Sigma^*, \resleq{f}{k})$ is a
            \kl{well-quasi-ordering};
        \item \label{n-poly-well:item} every $\prefleq$-increasing sequence
            of $\Sigma^*$  is a \kl{good sequence}
            for $\resleq{f}{k}$;
        \item \label{n-poly-residual:item} The
            \kl{$k$-residual transducer}
            of 
            $f$ exists.
    \end{enumerate}
    If $f$ is \kl{commutative}, the  
    properties are decidable, and the conversions effective.
\end{theorem}
\begin{proof}
    \cref{n-poly-1-transd:item} implies \cref{n-poly-k:item} by
    definition. Then,
    \cref{n-poly-k:item} implies \cref{n-poly-wqo:item} by
    \cref{n-poly-k-implies-wqo:lemma}.
    The implication \cref{n-poly-wqo:item} $\Rightarrow$ \cref{n-poly-well:item}
    is obvious.
    Then, \cref{wqo-implies-termination:lemma} proves
    that \cref{n-poly-well:item} implies \cref{n-poly-residual:item}.
    Furthermore, because a \kl{$k$-residual transducer} is a \kl{$\NPoly[k-1]$-transducer},
    \cref{n-poly-residual:item} implies \cref{n-poly-1-transd:item}.
    Finally, \cref{residual:algo}
    is effective as soon as $\resleq{f}{k}$ is decidable, which 
    is the case when $f$ is a \kl{commutative}
    \kl{$\Rel$-polyregular} function, as shown in
    \cite[Theorem 31]{LOPEZ24}.
\end{proof}

\section{Aperiodicity and Star-Free Functions}
\label{aperiodic-star-free:sec}

\AP In this section, we investigate the connection between \kl{star-free}
\kl{$\Nat$-polyregular} functions and structural properties of their
\kl{residual transducers}. This is motivated by the fact that in the case of
$\Rel$ outputs, belonging to $\ZSF[k]$ is equivalent to having a
\kl{$k$-residual transducer} without \kl{counters} and with $\ZSF[k-1]$ labels
on its transitions (see \cref{H-transducers:thm}).

\AP Unfortunately, the \kl{residual transducer} does not seem to provide any
characterization of \kl{star-free} \kl{$\Nat$-polyregular} functions. Indeed,
we can exhibit a function $f \in \NSF[1]$ such that its \kl{$1$-residual
transducer} contains a \kl{counter}, as shown in
\cref{non-aperiodic-residual-transd:ex}. Even more, we can prove
that any choice of \kl{$\NPoly[1]$-transducer} with at most $2$ states computing
our example function will contain a \kl{counter}.

\begin{example}
    \label{non-aperiodic-residual-transd:ex}
    Let us define
    $\BadExKo(\varepsilon) = 1$,
    $\BadExKo(a) = 0$,
    $\BadExKo(a^2) = 1$,
    and $\BadExKo(a^n) = n - 3$ for all $n \geq 3$.
    The \kl{$0$-residual transducer} of $\BadExKo$ has a \kl{counter} and two states,
    and is in fact a \kl{$\NSF[0]$-transducer}.
    Furthermore,
    every \kl{$\NPoly[0]$-transducer} with at most two states contains a \kl{counter}.
\end{example}
\begin{proof}
    It is quite clear that $\BadExKo$ cannot be computed by a \kl{$\NPoly[0]$-transducer}
    with one state, as $\Deriv{f}{a}{\varepsilon}$ takes negative values.
    Let us now consider a \kl{$\NPoly[0]$-transducer} $\aTransd$  with two states $q_0$ and $q_1$ without 
    \kl{counters},
    such that $q_0$ is the initial state, $\delta(q_0, a) = q_1$, 
    that computes $\BadExKo$.
    Because $\aTransd$ has no \kl{counters}, we conclude that $\delta(q_1,a) = q_1$.
    Then, because $\aTransd$ \kl{computes} $\BadExKo$, we must have
    $\aTransd(q_0, a) = \BadExKo(a) = 0 = \BadExKo(aaa) = \aTransd(q_0, aaa)$.
    As a consequence, 
    $F(q_1) = 0$, 
    $\lambda(q_1, a)(\varepsilon) = 0$,
    and
    $\lambda(q_1, a)(a) = 0$.
    However, this means 
    that $\aTransd(q_0, aa) = F(q_1) + \lambda(q_1, a)(a)  + \lambda(q_1, a)(\varepsilon) = 0$,
    which contradicts the fact that $\BadExKo(aa) = 1$.

    Let us now exhibit the \kl{$0$-residual transducer} of $\BadExKo$. It has
    two states $q_0$ (representing $\varepsilon$) and $q_1$ (representing $a$),
    with $q_0$ being the initial state. The map $F$ is defined by $F(q_0) = 1$
    and $F(q_1) = 0$. Then, $\delta(q_0, a) = q_1$, and $\delta(q_1, a) = q_0$.
    Finally, $\lambda(q_0, a)(w) = 0$ for all $w \in \Sigma^*$,
    and $\lambda(q_1, a)(w)$ is defined as $0$ if $\card{w} \leq 2$ and
    $2$ otherwise.
    This is correct because
    $\lambda(q_0, a) = \Deriv{f}{a}{a} = 0$
    and
    $\lambda(q_a, a) = \Deriv{f}{aa}{\varepsilon} = 2 \times \ind{\card{w} > 2}$.
    Notice that the defined function are \kl{star-free}, hence that
    the \kl{$0$-residual transducer} of $\BadExKo$ is an \kl{$\NSF[0]$-transducer}.
\end{proof}

In the specific case of $\NPoly[0]$, which corresponds to $\Nat$-linear
combination of indicator functions of regular languages, we can however recover
a characterization of \kl{star-free} functions in terms of the \kl{residual
transducer}.

\begin{lemma} 
    \label{aperiodic-iff-residual:lem}
    Let $f \in \NPoly[0]$. Then,
    $f \in \NSF$ if and only if 
    $f \in \ZSF$, 
    if and only if 
    the \kl{$0$-residual transducer} of $f$ is \kl{counter-free}.
\end{lemma}
\begin{proof}
    It is clear that $\NSF \subseteq \ZSF$. Furthermore, if the \kl{$0$-residual
    transducer} of $f$ is \kl{counter-free}, then $f \in \NSF$
    because $f$ can be computed as a $\Nat$-linear combination of
    indicator functions of star-free languages (computed by the automaton).

    Assume now that $f \in \ZSF$, and let us prove that the \kl{$0$-residual
    transducer} of $f$ is \kl{counter-free}. Note that because $f \in
    \NPoly[0]$, $u \resleq{f}{0} v$ if and only if $\app{f}{u} = \app{f}{v}$.
    In particular, in a \kl{$0$-residual transducer} of $f$, two states that
    represent the same \kl{residual} must be incomparable for the prefix relation.

    Let $(q,w^n)$ be a counter with $n \geq 1$. This means that $\delta(q, w^n)
    = q$ in the automaton, and implies that $q \resleq{f}{0} qw^n$, hence that
    $\app{f}{q} = \app{f}{qw^n}$. Because $f \in \ZSF$, we know that
    $\app{f}{qw^n} = \app{f}{qw^{n+1}}$, hence that $\app{f}{qw} = \app{f}{q}$.

    Let us write $t \defined \delta(q,w) = \delta(q,w^{n+1})$. We know that
    $\app{f}{q} = \app{f}{t}$. Assume by contradiction that $t$ and $q$ are
    incomparable for the prefix relation. Let us split $w = w_1 w_2$ where
    $w_1$ is the shortest prefix of $w$ such that $s_0 \defined \delta(q,w_1)$
    is an ancestor of $q$ and of $t$ for the prefix relation, it must exist
    because $\delta(q,w_1 w_2) = t$.

    Now, consider $s_1 \defined \delta(t, w_1)$. Assume by contradiction that
    $s_0$ is not comparable with $s_1$ for the prefix relation. Then, consider
    the smallest prefix $v$ of $w_1$ such that $\delta(t, v)$ is a strict
    prefix of $s_0$. It must exist, otherwise $s_0$ is always a prefix of
    $s_1$. Because $\app{f}{t} = \app{f}{q}$, we conclude that $\app{f}{tv} =
    \app{f}{qv}$. However, this contradicts the minimality of $w_1$, since
    $\delta(t,v)$ is an ancestor of $q$ and $t$.

    We have proven that $s_0$ and $s_1$ are comparable, hence they are equal,
    since $\app{f}{s_1} = \app{f}{t w_1} = \app{f}{q w_1} = \app{f}{s_0}$.
    Finally, we have proven that $\delta(q, w_1) = s_0$, $\delta(s_0, w_2) =
    t$, and $\delta(s_0, w_2) = \delta(t, w_1w_2) = q$ which is absurd.
    As a consequence $t$ and $q$ were comparable for the prefix relation,
    hence equal, and therefore $\delta(q, w) = q$.
\end{proof}

\AP In the hope of generalizing this result to higher growth rates, we
introduce the notion of \intro{aperiodic ordering} of $\Sigma^*$, designed to
mimic the notion of \kl{aperiodic monoid} in the context of regular
languages. Let us recall that a monoid $M$ is \intro(monoid){aperiodic} whenever for
all $x \in M$, there exists $n \in \Nat$ such that $x^n = x^{n+1}$. In the
specific case of finite monoids, this $n$ can be chosen uniformly for all
elements of the monoid. We therefore state that an ordered set $(\Sigma^*,
\leq)$ is \reintro[aperiodic ordering]{aperiodic} whenever for all $u, w \in
\Sigma^*$, there exists $N_0 \in \Nat$, such that the sequence $\seqof{uw^n}[n
\geq N_0]$ is non-decreasing. Finally, we introduce the \emph{star-free
variant} $(\intro*\resleqsf{f}{k})$ of $(\resleq{f}{k})$, defined by $u
\reintro*\resleqsf{f}{k} v$ whenever $\Deriv{f}{v}{u} \in \NSF[k-1]$.
With these definitions at hand, we are ready to state our main conjecture.

\begin{conjecture}
    \label{sf-no-periods-on-sequences:conj}
    For all $k \in \Nat$ and $f \colon \Sigma^* \to \Rel$,
    $f \in \NSF[k]$ if and only if $(\Sigma^*, \resleqsf{f}{k})$ is an \kl[aperiodic ordering]{aperiodic}
    \kl{well-quasi-ordering}.
\end{conjecture}

Note that \cref{sf-no-periods-on-sequences:conj} cannot rely on the current
definition of a \kl{residual transducer}, even when using its \emph{star-free}
variant (with $\NSF[k-1]$ labels), because of
\cref{non-aperiodic-residual-transd:ex}. However, we can already prove one
direction of this conjecture, namely that if $f \in \NSF[k]$, then $(\Sigma^*,
\resleqsf{f}{k})$ is an \kl(ordering){aperiodic} \kl{well-quasi-ordering}. This
relies on the characterization of $\NSF[k]$ as the set of functions that count
the number of realizations of some $\FO$ formulas, which gives us both the
\kl{well-quasi-ordering} property and the \kl[aperiodic ordering]{aperiodic}
property. To simplify reasoning, we will leverage results from \cite{LOPEZ24}
regarding \intro{commutative} \kl{star-free} \kl{$\Nat$-polyregular functions},
whereby \kl{commutative} means that the function is invariant under
permutations of its input word.

\begin{lemma}
    \label{sf-no-periods-on-sequences:lemma}
    Let $k \in \Nat$, and $f \in \NPoly[k]$. If $f \in \NSF[k]$, then
    $(\Sigma^*, \resleqsf{f}{k})$ is an
    \kl[aperiodic ordering]{aperiodic} \kl{well-quasi-ordering}.
\end{lemma}
\begin{proof}
    Let $f \in \NSF[k]$, $q \in \Nat$, and 
    write $f = \sum_{i=1}^n m_i \times \counting{\phi_i(\vec{x})}$, where
    $\phi_i \in \FO$ has quantifier rank at most $q$
    and $\card{\vec{x}} = k$.

    As in the proof of \cref{n-poly-k-implies-wqo:lemma}, we are going to
    assign a tuple of integers to a word $u \in \Sigma^*$ by counting the
    number of realizations of each $\FO^q$ type of at most $k$ variables in
    $u$. To that end, let us write $\FO^q(u)$ this vector.

    First, let us notice that for us to conclude, it suffices to prove that for
    some $n \in \Nat$, the inequality $\FO^q(uw^n) \leq \FO^q(uw^{n+1})$ holds,
    since $\Deriv{f}{uw^{n+1}}{uw^n} = \app{f}{uw^{n+1}} - \app{f}{uw^n}$ will
    be obtained as a non-negative combination of counting first-order types in
    the argument.

    Let $t$ be a first order type with at most $k$ free variables and
    quantifier rank at most $q$. The map $g_t \colon X \mapsto
    \counting{t}(uw^X)$ is a \kl{commutative} \kl{star-free $\Nat$-polyregular
    function}. As a consequence, there exists $N_0$ and a $K$ such that $g_t(X)
    = P(X)$ for all $X \geq N_0$, where $P(X+K) \in \Nat[X]$
    \cite{LOPEZ24}. As a consequence, $g_t(X + N_0 + K+1) - g_t(X + N_0 + K)
    \in \Nat$ for all $X \geq 0$.

    Because there are finitely many non-equivalent $\FO^q$ types with at most 
    $k$ free variables, we can take the maximum of the $K$'s obtained for each 
    of those, and conclude.
\end{proof}

\bibliographystyle{plainurl}

\end{document}